\documentclass[review]{elsarticle}
\usepackage{lineno,hyperref}
\usepackage{graphicx}
\usepackage{makecell}
\usepackage{caption}
\usepackage{booktabs}
\usepackage{amsthm,amsmath,amssymb}
\usepackage{mathrsfs}
\usepackage{array}
\biboptions{sort&compress}
\modulolinenumbers[5]

\journal{Journal of \LaTeX\ Templates}

\begin{document}
\captionsetup[figure]{labelfont={bf},labelformat={default},labelsep=period,name={Fig.}}

\begin{frontmatter}
\title{Personalized recommendation systems based on social relationships and historical behaviors}
\author[inst1]{Yan-Li Lee}
\author[inst2]{Tao Zhou\corref{cor1}}
\cortext[cor1]{Corresponding author.}
\ead{zhutou@ustc.edu}
\author[inst3]{Kexin Yang}
\author[inst1]{Yajun Du\corref{cor1}}
\ead{duyajun@mail.xhu.edu.cn}
\author[inst4]{Liming Pan}

\address[inst1]{School of Computer and Software Engineering, Xihua University, Chengdu 610039, China}
\address[inst2]{CompleX Lab, University of Electronic Science and Technology of China, Chengdu 611731, China}
\address[inst3]{College of Computer Science, Sichuan
University, Chengdu 610065, China}
\address[inst4]{School of Computer and Electronic Information, Nanjing Normal University, Nanjing 210023, China}

\begin{abstract}
Previous studies show that recommendation algorithms based on historical behaviors can provide satisfactory recommendation performance. Many of these algorithms pay attention to the interest of users, while ignore the influence of social relationships on user behaviors. Social relationships not only carry intrinsic information of similar consumption tastes or behaviors of users, but also imply the influence of individuals on their neighbors. In this paper, we assume that social relationships and historical behaviors are related to the same implicit factors. Based on this assumption, we propose an algorithm to jointly utilize social relationships and historical behaviors by the linear optimization technique. We test the performance of our algorithm for four types of users, including all users, active users, inactive users and cold-start users. Results show that the proposed algorithm outperforms benchmarks in four types of scenarios subject to recommendation accuracy and diversity metrics. We further design a randomization model to explore the contribution of social relationships to the recommendation performance, and the result shows that the contribution of social relationships depends on the coupling strength between social relationships and historical behaviors. 
\end{abstract}

\end{frontmatter}

\section{Introduction}
\label{introduction}
With the rapid development of Web 2.0, people enjoy the convenience it brings while also suffering from the information overload dilemma. We have to spend more and more time finding out the information that we are interested in. It not only degrades our surfing experience, but also reduces social productivity \cite{hemp2009death}. As a specialized tool, recommendation systems tackle this problem by personalized recommendation \cite{lu2012recommender,yu2016network,hui2022personalized,chiu2021developing,gao2020mining}, which employs users’ historical behaviors, personal profiles, the tag information and other relevant information to recommend objects to users. One of the keys to achieve good recommendation performance is to find out the interaction mechanism between users and objects as accurately as possible. If we can reveal mechanism intrinsic to interactions between users and objects, we can design a well-performing recommendation algorithm. Some people chooses one object because the object is recommended by his friend. However, Some people may unknowingly choose the same objects as strangers because they have the same taste. The latter is the underlying idea of the well-known collaborative filtering algorithms \cite{schafer2007collaborative,Koren2015advances}. That is, similar users tend to select similar objects. However, classical collaborative filtering algorithms ignore the influence of social relationships on users' historical behaviors, and it is practically impossible to obtain the social influence only by considering users' historical behaviors. This problem exists in classical physical dynamics algorithms as well \cite{zhou2007bipartite,zhang2007heat}.  

In recent years, the booming of social media makes it possible to acquire the social influence. The commercial value of social media has also became a new source of economic growth \cite{chen2014exploring,hutter2013impact,arazy2009improving}. Many studies \cite{hwang2010coauthorship, symeonidis2011product,zheng2021personalized,zhou2022point} show that social media can improve the recommendation performance. The underlying reasons are twofold: on the one hand, homophily theory \cite{mcpherson2001birds} indicates that users with social relationships tend to share similar consumer tastes or behaviors. For example, obese persons in a social network tend to gather together \cite{christakis2007spread}, and smokers or nonsmokers in a social network also tend to gather together \cite{christakis2008collective}. On the other hand, users' reviews and consumer opinions for objects in social media will influence their neighbors' purchase decisions and attitudes toward the objects. The phenomena has been revealed by extensive empirical studies: a user's purchase probability to a DVD increases with the number of incoming recommendations from his friends \cite{leskovec2007dynamics}; The communication between buyers in Taobao (a consumer marketplace in China) will largely drive the purchasing activity of users \cite{guo2011role}; Users with social relationships in an Asian mobile network tend to make similar purchase decisions in purchasing caller-back tones \cite{ma2015latent}; Connected users in Cyworld (an online social networking site in Korea) will be positively affected by their friends' purchase behaviors \cite{iyengar2009friends}; The purchase willingness of users in QQ (a social network in China) is more likely influenced by the number of prior adoptions in their neighborhoods than the well-connected neighbors \cite{zhu2016exploring}. In summary, social relationships are closely related to users' historical behaviors.

\begin{figure}[t]
\setlength{\abovecaptionskip}{0pt}
\centering
	\includegraphics[width=1\textwidth]{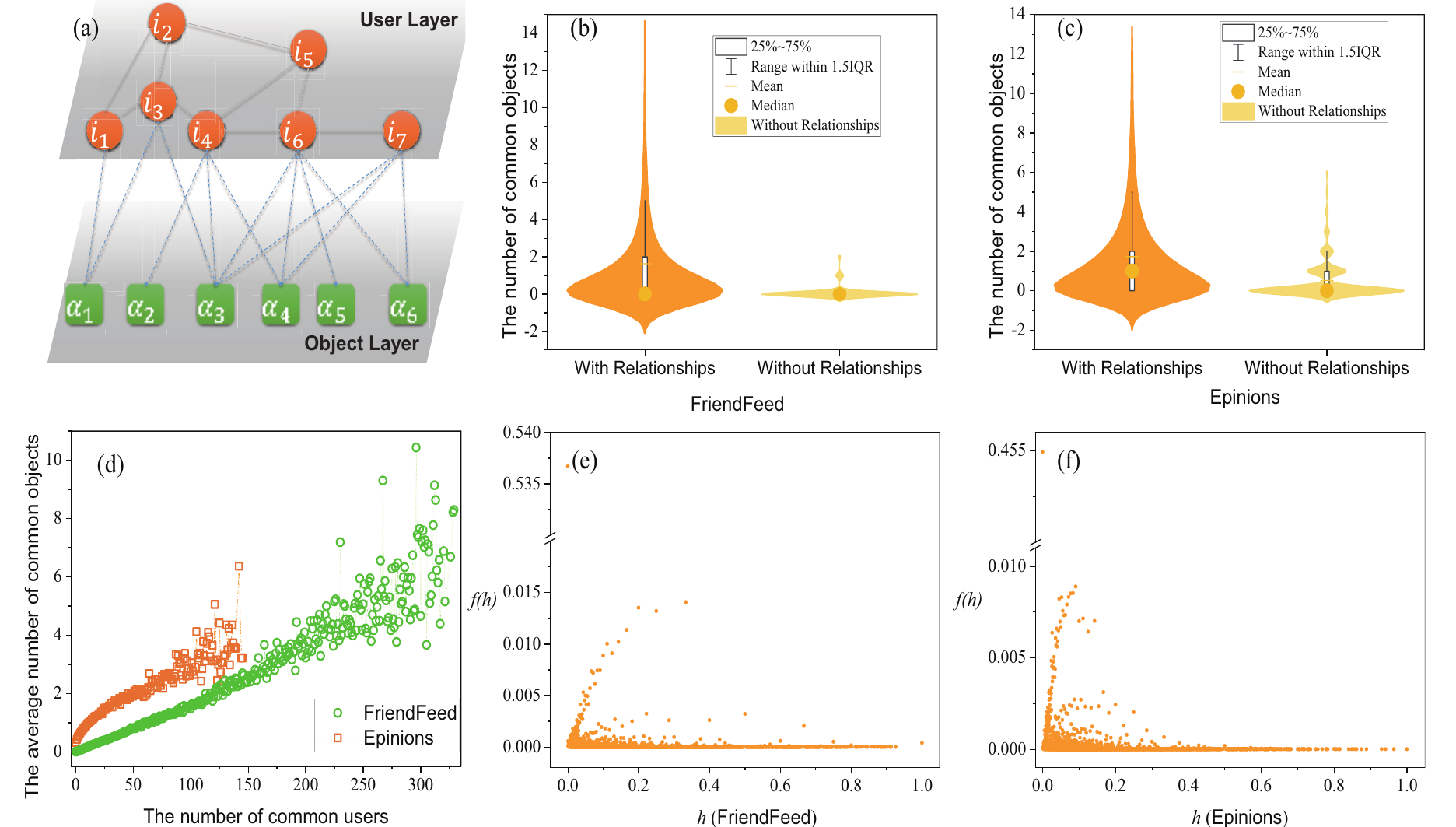}
    \caption{Social relationships vs. Historical behaviors. (a) The illustration of data structures of FriendFeed and Epinions. (b)-(c) The distribution of the number of common objects collected by users in the user-object interaction network. (d) The average number of common objects collected by users in the user-object interaction network vs. the number of common users shared by users in the social network. (e)-(f) The behavioral conversion rate $h$ of users, and $f(h)$ denotes the proportion of $h$.}
\label{fig.1}
\end{figure}

However, not all social relationships are relevant to users' historical behaviors. Here we take FriendFeed and Epinions as examples (see Section~\ref{datasets} for detailed descriptions of datasets). We show the corresponding data structure in Fig.~\ref{fig.1}(a). Namely, interactions take place among users in the user layer, and also take place between the user layer and the object layer. Overall speaking, users with direct social relationships tend to interact with more objects than users without direct social relationships (see Fig.~\ref{fig.1}(b) and Fig.~\ref{fig.1}(c)), and strongly similar users in social networks tend to interact with more objects than weakly similar users (see Fig.~\ref{fig.1}(d)). Taking a close look, one can find that behavioral conversion rates brought by social relationships vary widely. The behavioral conversion rate here is defined as the contribution from one user to its’ immediate neighbors. For example, in Fig.~\ref{fig.1}(a), objects $\alpha_2$, $\alpha_3$ and $\alpha_4$ are collected by user $i_4$, and objects $\alpha_3$, $\alpha_4$, $\alpha_5$ and $\alpha_6$ are collected by the neighbor $i_6$ of user $i_4$. Thus, the behavioral conversion rate from user $i_6$ to user $i_4$ is $h_{i_6i_4}=\frac{|\{\alpha_2,\alpha_3,\alpha_4\}\cap\{\alpha_3,\alpha_4,\alpha_5,\alpha_6\}|}{|\{\alpha_2,\alpha_3,\alpha_4\}|}=\frac{2}{3}$, and the behavioral conversion rate from user $i_4$ to user $i_6$ is $h_{i_4i_6}=\frac{|\{\alpha_2,\alpha_3,\alpha_4\}\cap\{\alpha_3,\alpha_4,\alpha_5,\alpha_6\}|}{|\{\alpha_3,\alpha_4,\alpha_5,\alpha_6\}|}=\frac{1}{2}$. We show the distribution of behavioral conversion rates of users on FriendFeed and Epinions in Fig.~\ref{fig.1}(e) and Fig.~\ref{fig.1}(f). One can find that, on FriendFeed and Epinions, social relationships with behavioral conversion rates larger than 0.2 separately account for 7.84\% and 2.12\%, and social relationships with behavioral conversion rates of 0 separately account for 53.67\% and 45.49\%. In other words, many of the explicit social relationships are not relevant to users' historical behaviors. This inspires us to pay much more attention to increase the strength of social relationships related to users' historical behaviors and decrease the strength of social relationships unrelated to or weakly related to users' historical behaviors.

To this end, we assume that social relationships and historical behaviors are related to the same implicit factors, thus contributions of the two types of relations can be jointly constrained by each other through implicit factors. We employ the linear optimization technique \cite{pech2019link} to obtain contribution weights of implicit factors through jointly constraints from social relationships and historical behaviors (i.e., user-object interactions), and then use the weights to make recommendations. We name this algorithm as SBLO. To test the performance of SBLO, we conduct experiments for four types of users including all users, active users (users with enough historical behaviors), inactive users (users with a small number of historical behaviors) and cold-start users (users without historical behaviors). Results show that SBLO outperforms 6 benchmarks subject to accuracy metrics, and SBLO greatly enhances the recommendation accuracy for inactive users and cold-start users. To our surprise, SBLO is competitive with algorithms dedicated to the diversity or accuracy-diversity dilemma.

\section{Related works}
The emergence of social media promotes social relationships to be widely used in recommendation systems. Many effective algorithms that consider social relationships are proposed including matrix factorization algorithms, probabilistic-based algorithms and others\cite{tang2013social, yang2014survey,yang2012bayesian,Hao2009learning,Hao2008Sorec,jiang2012social,yu2011adaptive,hong2016latent,kim2014twilite,xiao2022multi,zhou2022point,ahmadian2022reliable}. These earlier social recommendation algorithms need extra information (such as users’ rating records, users’ profiles, domain knowledge) to predict interactions between users and objects. However, the extra information not always be readily available. Previous studies show that algorithms based on topology information also can achieve significant recommendation performance \cite{chen2018trust, nie2014information, deng2017general}. This section mainly introduces some representative algorithms based on topology information. Specifically, we will first introduce recommendation algorithms that consider the topology information of user-object interaction networks, which can be constructed by historical behaviors. Then, we will introduce recommendation algorithms that consider the topology information of social networks and user-object interaction networks, which can be constructed by social relationships and historical behaviors, respectively.

Historical behaviors are used to achieve users’ interests or the relevance among objects. Collaborative filtering algorithms and physical dynamics models are two classical types of recommendation algorithms that consider historical behaviors. Collaborative filtering algorithms (CF) can be further grouped into memory-based CF algorithms \cite{rich1979user, sarwar2001item} and model-based CF algorithms \cite{breese1998empirical, shani2005mdp, chen2008group}. The former depends on the assumption that similar users tend to interact with similar objects; The latter utilize history behaviors to learn and infer users' consumer patterns. Physical dynamics algorithms mainly consider two different physical processes in a user-object interaction network, which are the mass diffusion process \cite{zhou2007bipartite} (MD, also named as probabilistic-spreading, ProbS) and the local heat conduction process \cite{zhang2007heat} (HC, also named as heat-spreading, HeatS). The mass diffusion process can generate highly accurate recommendations, and the local heat conduction process can generate highly diverse recommendations. Based on these two processes, many well-performing variants are proposed \cite{yu2016network,zhou2010solving, lu2011information, liu2011information, zhou2009accurate}. Many of them mainly focus on solving the accuracy-diversity dilemma through different perspectives, including the hybridization algorithm (HHP) \cite{zhou2010solving} that combines MD \cite{zhou2007bipartite} and HC \cite{zhang2007heat}, the preferential diffusion algorithm (PD) \cite{lu2011information} and the biased heat conduction algorithm (BHC) \cite{liu2011information}. Furthermore, the physical processes have been extended to networks with label information \cite{zhang2010personalized, zhang2010solving, zhang2011tag, xu2020recommending}. The above algorithms all can recommend objects that are difficult to perform content analysis like movies, music or artwork. Nevertheless, their recommendation accuracy is low when historical behaviors is insufficient, and even fail to handle new users or new objects.

Further consideration of social relationships makes the performance of collaborative filtering algorithms and physical dynamics models improved. Three typical ways are used to incorporate historical behaviors and social relationships under the collaborative filtering framework or the physical dynamics framework. The first way is to enhance similarities of users with explicit social relationships. More specifically, initial similarities are obtained based on user-object interaction networks, and similarities among users with explicit social relationships will be further enhanced. CosRA+T \cite{chen2018trust} is such an algorithm which uses the CosRA index \cite{chen2017vertex} to obtain users’ similarities. It can achieve significant improvement with a simple idea. The second way is to calculate similarities among users based on a user-object interaction network and a social network, respectively. Then, final similarities among users will be obtained by combining the two types of similarities. One of the representative algorithms is RWR-based algorithm \cite{nie2014information}, which proposes a nonlinear idea to combine the social similarity and the personal preference similarity. The last way is to integrate a user-object interaction network and a social network as one network, and perform a specific physical dynamics process in this network \cite{deng2017general}. The first two treatments can be applied to both the collaborative filtering framework and the physical dynamics framework.  

\section{Methods}
\label{methods}
In this section, we first describe the formalized process of the proposed algorithm SBLO, and then show the corresponding optimized process.

\subsection{Formalization}
\label{formalization}

\begin{figure}[t]
\setlength{\abovecaptionskip}{0pt}
\centering
	\includegraphics[width=0.6\textwidth]{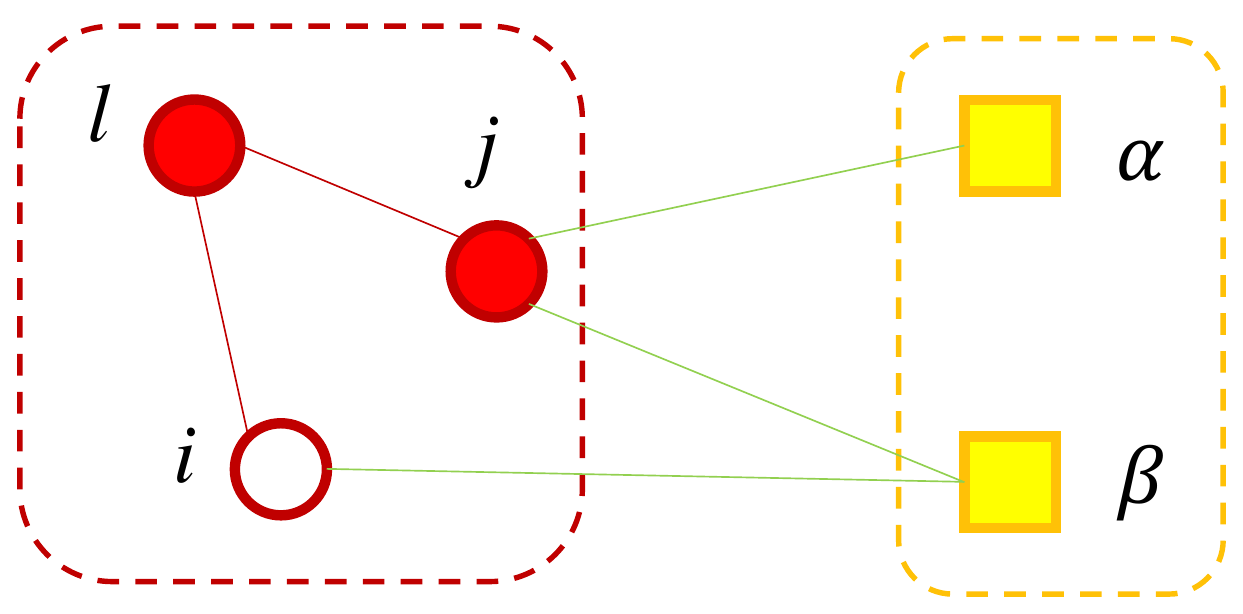}
    \caption{A toy model for the relevance between social relationships and interactions between users and objects.}
\label{fig.2}
\end{figure}

We assume that there exist some implicit factors which are related to social relationships and interactions between users and objects. The corresponding weights of implicit factors can be inferred from these two types of relations, so that we can make better use of these two types of relations. We illustrate this idea by a toy model. As shown in Fig.~\ref{fig.2}, if user $i$ can influence user $j$ or both of them have the same interests, user $i$ has a high probability of becoming friends with user $j$'s friend $l$. Meanwhile, there is a high probability for user $i$ to choose object $\alpha$ that is collected by user $j$. Implicit factors in this toy model include common interests, peer influence, and other unknown factors, and these factors are closely related to users' social relationships and behaviors. The challenge is how to obtain contribution weights of these implicit factors. Next, we will show a way to infer weights of implicit factors. 

Denote the explicit social relationship matrix as $\mathbf{A}\in \mathbb{R}^{m\times m}$, in which,
\begin{equation}
a_{il}=\begin{cases}
1, $\quad \textit{if there is a social relationship between user $i$ and user $l$};$\\
0, $\quad$ otherwise.
\end{cases}\\
\end{equation}
Denote the explicit interaction matrix as $\mathbf{B}\in \mathbb{R}^{m\times n}$, in which,
\begin{equation}
b_{i\beta}=\begin{cases}
1, $\quad$ $\textit{if there is an interaction between user $i$ and object $\beta$}$;\\
0, $\quad$ otherwise,
\end{cases}\\
\end{equation}
where $m$ and $n$ denote numbers of users and objects, respectively. Let $s$ denote the weight of implicit factors between two users. Then, the score $p_{il}$ of the existence of the social relationship between user $i$ and user $l$ can be denoted as
\begin{equation}
\label{eq.1}
p_{il} = \sum_j s_{ij} a_{jl},
\end{equation}
and the score $q_{i\alpha}$ of the existence of the interaction between user $i$ and object $\alpha$ can be denoted as
\begin{equation}
\label{eq.2}
q_{i\alpha} = \sum_j s_{ij} b_{j\alpha}.
\end{equation}
That means the social relationship between user $i$ and user $l$ can be unfolded by a linear summation of weights of implicit factors between user $i$ and $l$'s immediate neighbors in the social network; Meanwhile, the existence of a user-object interaction between user $i$ and object $\alpha$ also can be unfolded by a linear summation of weights of implicit factors between user $i$ and users who have collected object $\alpha$.

Clearly, Eq.~(\ref{eq.1}) and Eq.~(\ref{eq.2}) can be rewritten in matrix forms as
\begin{equation}
\label{eq.3}
\mathbf{P} = \mathbf{S}\mathbf{A},
\end{equation}
and
\begin{equation}
\label{eq.4}
\mathbf{Q} = \mathbf{S}\mathbf{B},
\end{equation}
where $\mathbf{S}$ is an unknown implicit factor matrix. The core of this problem turns into how to solve the unknown matrix $\mathbf{S}$.

\subsection{Optimization}
\label{optimization}
To estimate the unknown matrix $\mathbf{S}$, we formulate it as a linear optimization problem
\begin{alignat}{2}
\min_{\mathbf{S}}\quad & \vert\vert 
\mathbf{S} \vert\vert_F^2 + \lambda_1\vert\vert \mathbf{A}-\mathbf{SA} \vert\vert_F^2 + \lambda_2\vert\vert \mathbf{B}-\mathbf{SB} \vert\vert_F^2.
\label{eq.5}
\end{alignat}
That is, the optimal matrix $\mathbf{S}$ needs to be able to fit $\mathbf{A}$ and $\mathbf{B}$ well at the same time. Then, we have 
\begin{alignat}{2}
F(\mathbf{S}) & = \vert\vert 
\mathbf{S} \vert\vert_F^2 + \lambda_1\vert\vert \mathbf{A}-\mathbf{S}\mathbf{A} \vert\vert_F^2 + \lambda_2\vert\vert \mathbf{B}-\mathbf{S}\mathbf{B} \vert\vert_F^2 \label{eq.6}\\ 
& = Tr(\mathbf{S}^T\mathbf{S})+\lambda_1 Tr((\mathbf{A}-\mathbf{S}\mathbf{A})^T(\mathbf{A}-\mathbf{S}\mathbf{A})) + \lambda_2 Tr((\mathbf{B}-\mathbf{S}\mathbf{B})^T(\mathbf{B}-\mathbf{S}\mathbf{B})).\nonumber
\end{alignat}
The core to estimate $\mathbf{S}$ is to calculate the gradient of $F(\mathbf{S})$. Let 
\begin{equation}
\label{eq.7}
\nabla F(\mathbf{S}) = \mathbf{S}(\lambda_1\mathbf{A}\mathbf{A}^T + \lambda_2 \mathbf{B}\mathbf{B}^T)-(\lambda_1\mathbf{A}\mathbf{A}^T + \lambda_2 \mathbf{B}\mathbf{B}^T) + \mathbf{S} = 0.
\end{equation}
We can obtain the optimal solution of $\mathbf{S}$ as
\begin{equation}
\label{eq.8}
\mathbf{S}^* = (\lambda_1 \mathbf{A}\mathbf{A}^T + \lambda_2 \mathbf{B}\mathbf{B}^T)(\lambda_1 \mathbf{A}\mathbf{A}^T + \lambda_2 \mathbf{B}\mathbf{B}^T+\mathbf{I})^{-1},
\end{equation}
where $\mathbf{I}$ is the identity matrix. Accordingly, the scoring matrix of users on objects is
\begin{equation}
\label{eq.9}
\mathbf{R} = \mathbf{S}^*\mathbf{B}.
\end{equation}
The larger the element in $\mathbf{R}$, the higher the probability of one user to select the corresponding object. As a result, we obtain the proposed algorithm SBLO. 

\section{Experiments}
\label{experiments}
\subsection{Datasets}
\label{datasets}
\begin{table}[h!]
\centering
\footnotesize
\caption{Structural statistics of FriendFeed and Epinions. $m$ and $n$ are numbers of users and objects, respectively. $\vert E^A \vert$ and $\vert E^B \vert$ are numbers of social relationships in social networks and interactions in user-object interaction networks, respectively. $\langle k^A \rangle$ and $\langle k^B \rangle$ are average user degrees in social networks and user-object interaction networks, respectively. Sparsity(A)=$\frac{2\vert E^A \vert}{m^2}$ and Sparsity(B)=$\frac{\vert E^B \vert}{mn}$ denote the data sparsity of social networks and user-object interaction networks, respectively.}
\begin{tabular*}{\hsize}{lcccccccc}
\toprule
Datasets & $m$ & $n$ & $\vert E^A \vert$ & $\vert E^B \vert$ & $\langle k^A \rangle$ & $\langle k^B \rangle$ & $Sparsity(A)$ & $Sparsity(B)$\\
\midrule
FriendFeed&4148&5700&265497&96942&128&23&3$\times 10^{-2}$&4$\times 10^{-3}$\\
Epinions &4066&7649&167717&154122&82&37&2$\times 10^{-2}$&5$\times 10^{-3}$\\
\bottomrule
\end{tabular*}
\label{tb1}
\end{table}

FriendFeed and Epinions \cite{nie2014information} are used to test the performance of SBLO. FriendFeed (http://www.friendfeed.com/) is a real-time feed aggregation website that updates the information from media like Twitter, YouTube and Delicious, which contains follower-followee relationships and rating relationships between users and objects. Epinions (http://www.epinions.com/) is a consumer review website, which contains trust relationships and rating relationships between users and objects. Social networks are constructed based on follower-followee relationships or trust relationships. There is an unweighted and undirected link between user $i$ and user $j$ if there is at least one following relationship or one trust relationship between them. User-object interaction networks are constructed based on rating relationships. There is an unweighted and undirected link between user $i$ and object $\beta$, if the rating value from user $i$ to object $\beta$ is no less than 3 (both datasets with a 5-point rating scale from 1 to 5). Elementary statistics are shown in Table~\ref{tb1}. 

\begin{figure}[t]
\setlength{\abovecaptionskip}{0pt}
\centering
	\includegraphics[width=1\textwidth]{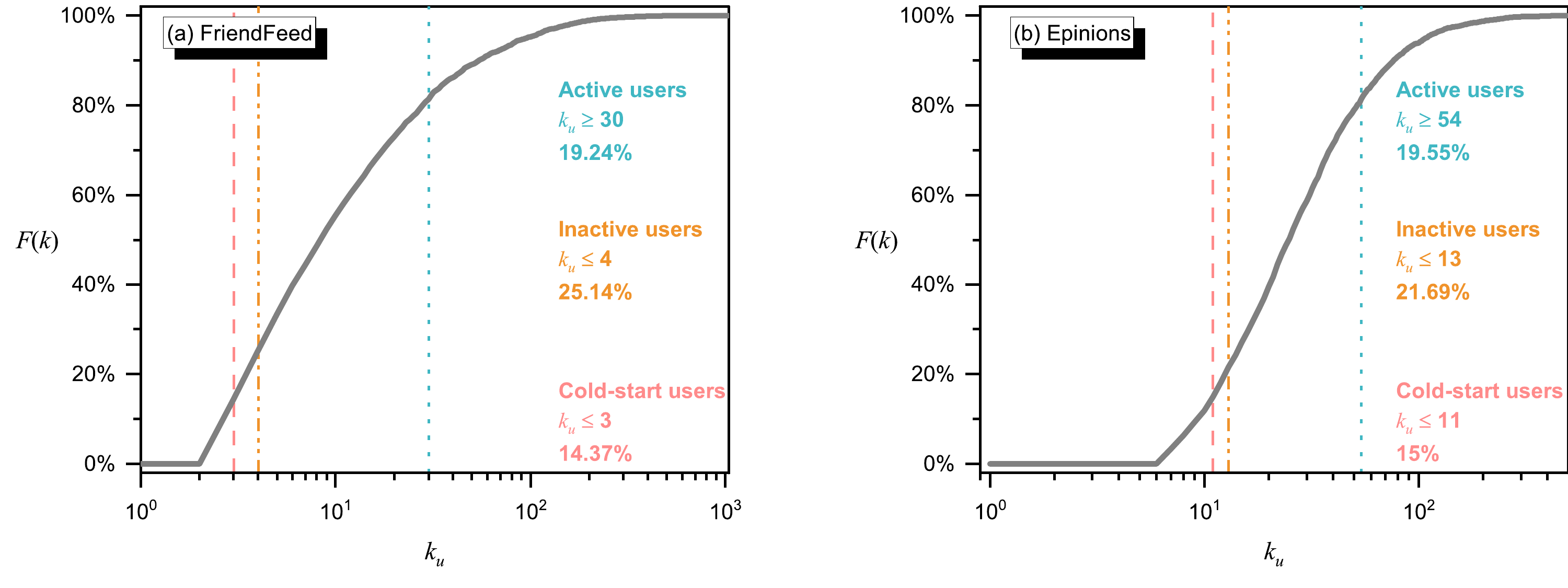}
    \caption{Cumulative probability distributions of user degrees in user-object interaction networks.}
\label{fig.3}
\end{figure}

To facilitate testing the performance of algorithms against data sparsity, we divide system users into active users, inactive users and cold-start users based on cumulative probability distributions of user degrees in user-object interaction networks (see Fig.~\ref{fig.3}(a)--Fig.~\ref{fig.3}(b)). On FriendFeed, $14.37\%$ users with user degrees no larger than 3 are regarded as cold-start users, $25.14\%$ users with user degrees no larger than 4 are regarded as inactive users, and $19.24\%$ users with user degrees no less than 30 are regarded as active users. Similarly, on Epinions, $15\%$ users with user degrees no larger than 11 are regarded as cold-start users, $21.69\%$ users with user degrees no larger than 13 are regarded as inactive users, and $19.55\%$ users with user degrees no less than 54 are regarded as active users.

\subsection{Metrics}
\label{metrics}
A good recommendation list is expected to have a high hit ratio and contain diverse information. Accordingly, accuracy metrics \cite{yang2015evaluating, shani2011evaluating}(precision, recall, F-score and AUPR) and diversity metrics (intra-similarity \cite{zhou2009accurate}, hamming distance \cite{zhou2008effect} and popularity) are introduced to evaluate the algorithmic performance. If $L_r$ objects among the top-$L$ selected objects are correctly recommended to the target user $i$, the corresponding precision value is 
\begin{equation}
\label{eq.10}
    Pre_i(L) = \frac{L_r}{L},
\end{equation}
and the corresponding recall value is
\begin{equation}
\label{eq.11} 
    Rec_i(L) = \frac{L_r}{\vert E_i^P \vert},
\end{equation}
where $\vert E_i^P \vert$ is the total number of relevant objects of user $i$. F-score ($F$) of user $i$ is defined as
\begin{equation}
\label{eq.12}
    F_i(L) = \frac{2Pre_i(L)Rec_i(L)}{Pre_i(L)+Rec_i(L)}.
\end{equation}
AUPR \cite{yang2015evaluating} of user $i$ is the area under the precision-recall curve, which is more suitable for Top-K recommendations. Compared to precision and recall, AUPR does not depend on the choice of $L$ and shows a more comprehensive evaluation by considering a range of $L$. Intra-similarity \cite{zhou2009accurate} ($I$) measures the recommendation diversity for a single user, which is defined as
\begin{equation}
\label{eq.13}
I_i(L)=\frac{1}{L(L-1)}\sum_{
\alpha \neq \beta \atop
\alpha, \beta \in O_i(L)}z_{\alpha\beta},
\end{equation} 
where $O_i(L)$ is the set of top-$L$ selected objects of user $i$. $z_{\alpha\beta}$ is the similarity of objects $\alpha$ and object $\beta$, which is calculated by the Salton similarity \cite{salton1983j} based on the user-object interaction network. Hamming distance \cite{zhou2008effect} ($H$) measures the recommendation diversity for different users, which is defined as
\begin{equation}
\label{eq.14}
H_i(L) = 1-\frac{C_{ij}(L)}{L},
\end{equation}
where $C_{ij}(L) = |O_i(L) \cap O_j(L)|$. Popularity ($Pop$) measures the novelty of the recommendation as
\begin{equation}
\label{eq.15}
Pop_i(L)=\frac{\sum_{\alpha\in o_i(L)}k_\alpha^B}{L},
\end{equation}
where $k_\alpha^B$ is the object degree of object $\alpha$ in user-object interaction networks. Finally, we obtain the system precision, recall, F-score, AUPR, intra-similarity and popularity by averaging over the corresponding individual performance for all related users. The system hamming distance is obtained by averaging over hamming distance values over all pairs of related users.

\subsection{Benchmarks}
\label{benchmarks}
We compare SBLO with 6 benchmarks, including three algorithms only based on user-object interaction networks (MD \cite{zhou2007bipartite}, HHP \cite{zhou2010solving} and PD \cite{lu2011information}), and three recommendation algorithms based on both user-object interaction networks and social networks (RWR-based algorithm \cite{nie2014information}, CosRA+T \cite{chen2018trust} and SocMD \cite{deng2017general}). In this subsection, for MD \cite{zhou2007bipartite}, HHP \cite{zhou2010solving} and PD \cite{lu2011information}, we use $k$ to denote the degree of each node in user-object interaction networks. For RWR-based algorithm \cite{nie2014information}, CosRA+T \cite{chen2018trust} and SocMD \cite{deng2017general}, we use $k^A$ to denote the degree of each node in social networks, and use $k^B$ to denote the degree of each node in user-object interaction networks.

MD \cite{zhou2007bipartite} determines the interest of the target user $i$ on object $\alpha$ by a resource-allocation process. Specifically, each object collected by the target user $i$ is allocated one unit resource. The resource will be equally distributed to all users who collect the object, and then the users will send back what they have received equally to all objects which they have collected. The resource distributed from object $\beta$ to object $\alpha$ is denoted as $W_{\alpha\beta}^{MD}$, which is
\begin{equation}
\label{eq.16}
W_{\alpha\beta}^{MD} =
\frac{1}{k_\beta}\sum_{l=1}^m\frac{b_{l\alpha}b_{l\beta}}{k_l}.
\end{equation}
Accordingly, the interest of user $i$  on object $\alpha$ can be written as $f_{i\alpha}^{MD}=\sum_{\beta=1}^n b_{i\beta}W_{\alpha\beta}^{MD}$.

HHP \cite{zhou2010solving} incorporates MD and HC by a hybridization parameter $\lambda$ to make a trade off between accuracy and diversity, and
\begin{equation}
\label{eq.17}
W_{\alpha\beta}^{HHP} = \frac{1}{k_\alpha^{1-\lambda}k_\beta^{\lambda}}\sum_{l\in\Gamma_\beta}\frac{b_{l\alpha}b_{l\beta}}{k_l},
\end{equation}
where $\Gamma_{\beta}$ is the set of users that collect the object $\beta$. Similarly, the interest of user $i$ on object $\alpha$ can be obtained following MD. 

To further improve the diversity and novelty of recommendations, at the last step of MD, PD \cite{lu2011information} makes the resource that object $\alpha$ receives from object $\beta$ proportional to $k_\alpha^\varepsilon$, hence
\begin{equation}
\label{eq.18}
W_{\alpha\beta}^{PD} =
\frac{1}{k_\beta k_\alpha^{-\varepsilon}}\sum_{l=1}^m\frac{b_{l\alpha}b_{l\beta}}{\mathcal{M}},
\end{equation}
where $\mathcal{M}=\sum_{r=1}^n a_{lr}k_r^\varepsilon$, and thereby $f_{i\alpha}^{PD}=\sum_{\beta=1}^n b_{i\beta}W_{\alpha\beta}^{PD}$.

The RWR-based algorithm \cite{nie2014information} independently obtains the social similarity $r_{ij}^A$ and the personalized preference similarity $r_{ij}^B$ based on the social network and the user-object interaction network, respectively, and thereby the user similarity is
\begin{equation}
\label{eq.19}
W_{ij}^{RWR-based} =(r_{ij}^A)^{\theta_1}(r_{ij}^B)^{\theta_2}.
\end{equation}
$r_{ij}^A$ is an element of $\vec{r}_i^A$, which is obtained based on a Random Walk with Restart (RWR) method \cite{tong2006fast} in a social network. Denote $\mathbf{T}$ as the transition matrix of a social network, where $T_{ij}=1/k_i^A$. The peer-to-peer influence $\vec{r}_i^A$ of user $i$ on other users can be calculated by $r_{i}^A=(1-\theta_3)(1-\theta_3\mathbf{T})^{-1}\vec{e_i}$, where $\vec{e_i}$ is a unit vector. $r_{ij}^B$ is calculated by the Salton similarity \cite{salton1983j} based on the user-object interaction network. $\theta_1$, $\theta_2$ and $\theta_3$ are three tunable parameters. Accordingly, $f_{i\alpha}^{RWR-based}=\sum_{j=1}^mW_{ij}^{RWR-based}b_{j\alpha}$.

CosRA+T \cite{chen2018trust} directly utilizes social relationships to control the resource that user $j$ receives from the target user $i$. Denote the original resource that user $j$ receives from user $i$ as $t_{ij} = \sum_{\beta=1}^n \frac{b_{i\beta}b_{j\beta}}{\sqrt{k_j^Bk_{\beta}^B}}$. The final resource that user $j$ receives from the target user $i$ is $t_{ij}^\theta$ if $a_{ij}=1$ and $t_{ij}$, otherwise. Mathematically, the interest of user $i$ on object $\alpha$ can be written as

\begin{equation}
\label{eq.20}
f_{i\alpha}^{CosRA+T}=\sum_{j=1}^m\frac{b_{j\alpha}}{\sqrt{k_j^Bk_\alpha^B}}(a_{ij}t_{ij}^\theta+(1-a_{ij})t_{ij}),
\end{equation}
where $\theta$ is a tunable parameter. 

SocMD \cite{deng2017general} introduces the mass diffusion process in a network that integrates the social network and the user-object interaction network, with the probability $p$ to diffuse in the user-object interaction network, and the probability $1-p$ to diffuse in the social network. The interest of user $i$ on object $\alpha$ is denoted as
\begin{equation}
\label{eq.21}
f_{i\alpha}^{SocMD}=p\sum_{l=1}^m\sum_{\beta=1}^n\frac{b_{l\alpha}b_{l\beta}b_{i\beta}}{k_l^Bk_{\beta}^Bk_i^B}+(1-p)\sum_{l=1}^m\sum_{j=1}^m\frac{b_{l\alpha}a_{lj}a_{ji}}{k_l^Bk_j^Ak_i^A}).
\end{equation}

\subsection{Results}
\label{results}
Let $E^A$ and $E^B$ denote sets of social relationships and user-object interactions, respectively. To test the algorithmic performance, we randomly divide $E^B$ into two parts: the training set $E^T$ contains $90\%$ of all interactions, and the remaining $10\%$ interactions constitute the probe set $E^P$. Obviously, $E^T\cap E^P = \emptyset$ and $E^T\cup E^P = E^B$. The considered recommendation algorithms will generate a recommendation list for each target user based on $E^T$ and $E^A$. We separately compare the 7 algorithms for four types of users, including all users, active users, inactive users and cold-start users. 

As shown in Table~\ref{tb2}, for all users, SBLO performs best subject to AUPR, precision, recall and F-score, and is competitive with CosRA+T in hamming distance and popularity. Many of algorithms that consider both historical behaviors and social relationships perform worse in intra-similarity. In other words, their recommendation lists for a single user are relatively homogeneous. This may be because recommendation probabilities of objects collected by the same social community are increased when social relationships are considered. While CosRA+T works well in intra-similarity since niche or unpopular objects are recommended in large probability.

\begin{table}
\centering
\footnotesize
\caption{Recommendation performance of 7 considered algorithms for all users on FriendFeed (top half) and Epinions (bottom half). Each result is averaged over 20 independent runs with the recommendation length $L$=50. Parameters of HHP, PD, CosRA+T, SocMD, RWR-based and SBLO are tuned to their optimal values subject to maximal AUPR. The best-performed results are emphasized in bold.}
\begin{tabular}{p{2cm}p{1.2cm}p{1.2cm}p{1.2cm}p{1cm}p{1cm}p{1cm}p{0.5cm}}
\toprule
Algorithms & AUPR & Pre & Rec & F & I & H & Pop\\
\midrule
MD&0.0204&0.0142&0.2355&0.0268&0.0968&0.9105&49 \\
HHP&0.0210&0.0146&0.2390&0.0275&0.0920&0.9354&43 \\
PD&0.0204&0.0145&0.2302&0.0272&\textbf{0.0870}&0.9515&39 \\
CosRA+T&0.0224&0.0155&0.2444&0.0291&0.0947&\textbf{0.9699}&\textbf{31} \\
SocMD&0.0204&0.0142&0.2355&0.0268&0.0968&0.9105&49 \\
RWR-based&0.0220&0.0150&0.2499&0.0282&0.1154&0.9178&47 \\
SBLO&\textbf{0.0239}&\textbf{0.0162}&\textbf{0.2682}&\textbf{0.0305}&0.1047&0.9561&39 \\
\midrule
MD&0.0170&0.0149&0.1782&0.0275&0.0885&0.6627&170 \\
HHP&0.0209&0.0171&0.1986&0.0314&0.0811&0.8453&122 \\
PD&0.0197&0.0164&0.1878&0.0302&\textbf{0.0746}&0.8508&121 \\
CosRA+T&0.0213&0.0175&0.2030&0.0323&0.0790&0.8849&107 \\
SocMD&0.0170&0.0149&0.1782&0.0275&0.0885&0.6627&170 \\
RWR-based&0.0184&0.0153&0.1779&0.0282&0.0911&0.7831&144 \\
SBLO&\textbf{0.0220}&\textbf{0.0179}&\textbf{0.2059}&\textbf{0.0329}&0.0816&\textbf{0.9134}&\textbf{102} \\
\bottomrule
\end{tabular}
\label{tb2}
\end{table}

\begin{table}
\centering
\footnotesize
\caption{Recommendation performance of 7 considered algorithms for inactive users on FriendFeed (top half) and Epinions (bottom half). Each result is averaged over 20 independent runs with the recommendation length $L$=50. Parameters of HHP, PD, CosRA+T, SocMD, RWR-based and SBLO are tuned to their optimal values subject to maximal AUPR. The best-performed results are emphasized in bold.}
\begin{tabular}{p{2cm}p{1.2cm}p{1.2cm}p{1.2cm}p{1cm}p{1cm}p{1cm}p{0.5cm}}
\toprule
Algorithms & AUPR & Pre & Rec & F & I & H & Pop\\
\midrule
MD&0.0177&0.0072&0.2701&0.0140&0.0966&0.9102&49 \\
HHP&0.0177&0.0073&0.2742&0.0141&0.0964&0.9130&49 \\
PD&0.0176&0.0072&0.2676&0.0140&0.0937&0.9277&45 \\
CosRA+T&0.0179&0.0072&0.2667&0.0139&\textbf{0.0931}&\textbf{0.9696}&\textbf{31} \\
SocMD&0.0187&0.0079&0.2996&0.0154&0.0983&0.8272&58 \\
RWR-based&0.0197&0.0081&0.3024&0.0157&0.1282&0.8752&53 \\
SBLO&\textbf{0.0218}&\textbf{0.0089}&\textbf{0.3330}&\textbf{0.0173}&0.1119&0.9281&46 \\
\midrule
MD&0.0136&0.0080&0.1841&0.0153&0.0886&0.6626&170 \\
HHP&0.0151&0.0087&0.1983&0.0167&0.0827&0.8256&129 \\
PD&0.0145&0.0082&0.1883&0.0157&0.0803&0.7915&140 \\
CosRA+T&0.0149&0.0085&0.1917&0.0162&\textbf{0.0786}&\textbf{0.8730}&\textbf{112} \\
SocMD&0.0138&0.0081&0.1828&0.0155&0.0893&0.6234&176 \\
RWR-based&0.0144&0.0083&0.1907&0.0160&0.1008&0.7174&159 \\
SBLO&\textbf{0.0160}&\textbf{0.0090}&\textbf{0.2069}&\textbf{0.0173}&0.0912&0.8575&126 \\
\bottomrule
\end{tabular}
\label{tb3}
\end{table}

\begin{table}[h]
\centering
\footnotesize
\caption{Recommendation performance of 7 considered algorithms for active users on FriendFeed (top half) and Epinions (bottom half). Each result is averaged over 20 independent runs with the recommendation length $L$=50. Parameters of HHP, PD, CosRA+T, SocMD, RWR-based and SBLO are tuned to their optimal values subject to maximal AUPR. The best-performed results are emphasized in bold.}
\begin{tabular}{p{2cm}p{1.2cm}p{1.2cm}p{1.2cm}p{1cm}p{1cm}p{1cm}p{0.5cm}}
\toprule
Algorithms & AUPR & Pre & Rec & F & I & H & Pop\\
\midrule
MD&0.0260&0.0283&0.1754&0.0488&0.0970&0.9108&49 \\
HHP&0.0298&0.0313&0.1906&0.0537&\textbf{0.0768}&\textbf{0.9723}&\textbf{30} \\
PD&0.0297&0.0311&0.1909&0.0535&0.0794&0.9692&32 \\
CosRA+T&\textbf{0.0314}&0.0325&\textbf{0.2005}&0.0559&0.0901&0.9690&32 \\
SocMD&0.0260&0.0283&0.1754&0.0488&0.0970&0.9108&49 \\
RWR-based&0.0287&0.0306&0.1882&0.0526&0.1093&0.9375&43 \\
SBLO&0.0312&\textbf{0.0329}&0.1996&\textbf{0.0565}&0.1056&0.9551&39 \\
\midrule
MD&0.0270&0.0340&0.1617&0.0562&0.0884&0.6626&170 \\
HHP&0.0369&0.0421&0.1979&0.0695&0.0717&0.9196&90 \\
PD&0.0364&0.0418&0.1945&0.0688&\textbf{0.0628}&0.9281&\textbf{87} \\
CosRA+T&0.0370&0.0418&0.1977&0.0690&0.0788&0.8876&106 \\
SocMD&0.0270&0.0340&0.1617&0.0562&0.0884&0.6626&170 \\
RWR-based&0.0328&0.0385&0.1816&0.0636&0.0923&0.8041&139 \\
SBLO&\textbf{0.0381}&\textbf{0.0434}&\textbf{0.2000}&\textbf{0.0713}&0.0760&\textbf{0.9383}&88 \\
\bottomrule
\end{tabular}
\label{tb4}
\end{table}

\begin{table}
\centering
\footnotesize
\caption{Recommendation performance of 4 considered algorithms for cold-start users on FriendFeed (top half) and Epinions (bottom half). Parameters of SocMD, RWR-based and SBLO are tuned to their optimal values subject to maximal AUPR. The best-performed results are emphasized in bold.}
\begin{tabular}{p{2cm}p{1.2cm}p{1.2cm}p{1.2cm}p{1cm}p{1cm}p{1cm}p{0.5cm}}
\toprule
Algorithms & AUPR & Pre & Rec & F & I & H & Pop\\
\midrule
SocMD&0.0243&0.0167&0.2780&0.0315&0.1098&0.7634&66 \\
RWR-based&0.0254&0.0163&0.2718&0.0308&0.1281&0.7426&75 \\
GRM&0.0087&0.0055&0.0923&0.0104&\textbf{0.0875}&0&112 \\
SBLO&\textbf{0.0337}&\textbf{0.0197}&\textbf{0.3277}&\textbf{0.0371}&0.0910&\textbf{0.9265}&\textbf{46} \\
\midrule
SocMD&0.0212&0.0263&0.1316&0.0439&0.1052&0.4694&201 \\
RWR-based&0.0187&0.0238&0.1187&0.0397&0.1134&0.2927&221 \\
GRM&0.0139&0.0188&0.0930&0.0312&0.1151&0&230 \\
SBLO&\textbf{0.0229}&\textbf{0.0289}&\textbf{0.1446}&\textbf{0.0482}&\textbf{0.1050}&\textbf{0.7317}&\textbf{163} \\
\bottomrule
\end{tabular}
\label{tb5}
\end{table}

It is a challenge to recommend suitable objects to inactive users in recommendation systems. Because algorithms can not accurately capture users' preferences with a small number of historical behaviors. We compare the performance of seven considered algorithms for inactive users in Table~\ref{tb3}. Again, SBLO performs best subject to AUPR, precision, recall and F-score, and is the runner-up subject to hamming distance and popularity. In addition, we also test the performance of considered algorithms for active users in Table~\ref{tb4}. To our surprise, SBLO still performs overall best in accuracy metrics, but performs slightly worse than algorithms including HHP and PD subject to hamming distance and popularity. We further test the algorithmic performance in an extreme case, i.e., recommendation for cold-start users. Cold-start users have no historical behaviors. In order to simulate this situation with existing datasets, we remove all user-object interactions of cold-start users. This part of interactions constitute the probe set $E^P$. Clearly, many personalized recommendation algorithms like MD, HHP, PD, CosRA+T do not work in this case. Comparatively speaking, the global ranking method (GRM), which ranks all objects in the descending order of object degrees, still works. The recommendation performance of GRM and three algorithms that consider both historical behaviors and social relationships is reported in Table~\ref{tb5}. Compared to benchmarks, SBLO performs spectacularly well with both higher accuracy and higher diversity.

\begin{figure}[h]
\setlength{\abovecaptionskip}{0pt}
\centering
	\includegraphics[width=1\textwidth]{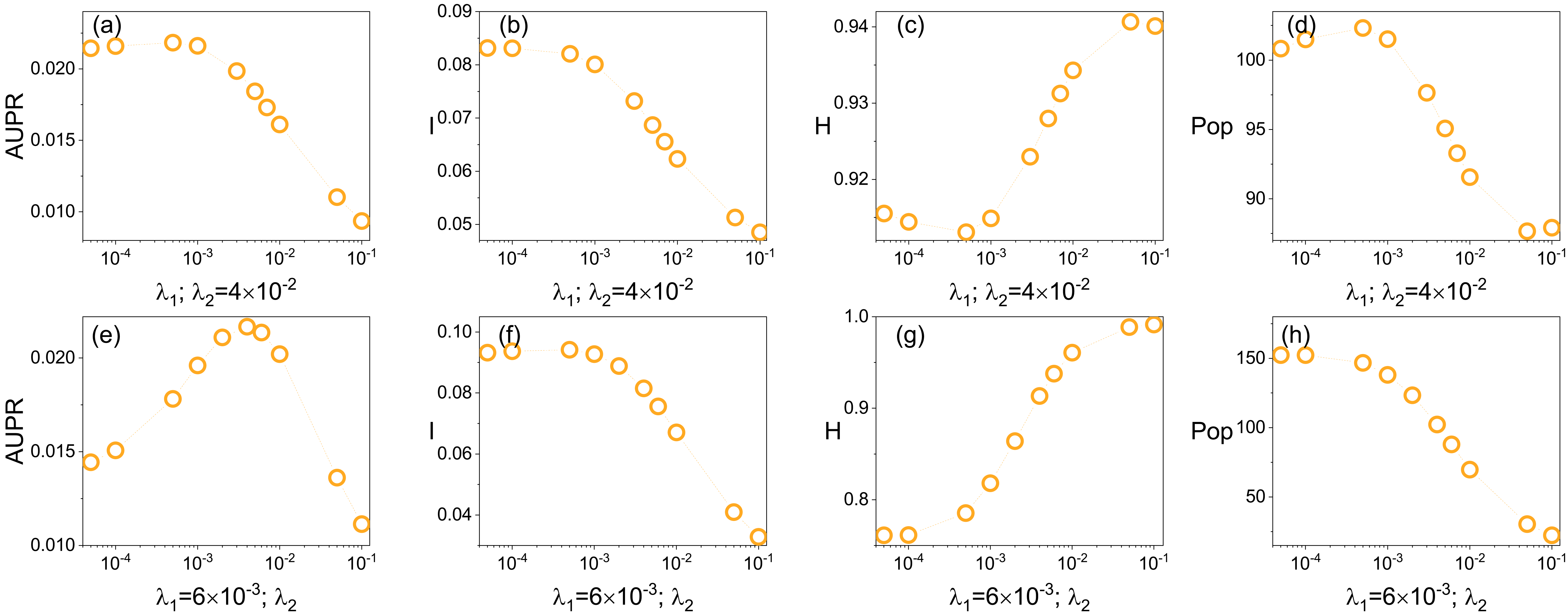}
    \caption{On Epinions,the performance of SBLO for all users, and all results are averaged over 20 independent runs. (a)-(d) AUPR vs. $\lambda_1$, $\lambda_2$=4$\times$ 10$^{-2}$. (e)-(h) AUPR vs. $\lambda_2$, $\lambda_1$=6$\times$ 10$^{-3}$.}
\label{fig.4}
\end{figure}

\begin{figure}[h]
\setlength{\abovecaptionskip}{0pt}
\centering
	\includegraphics[width=1\textwidth]{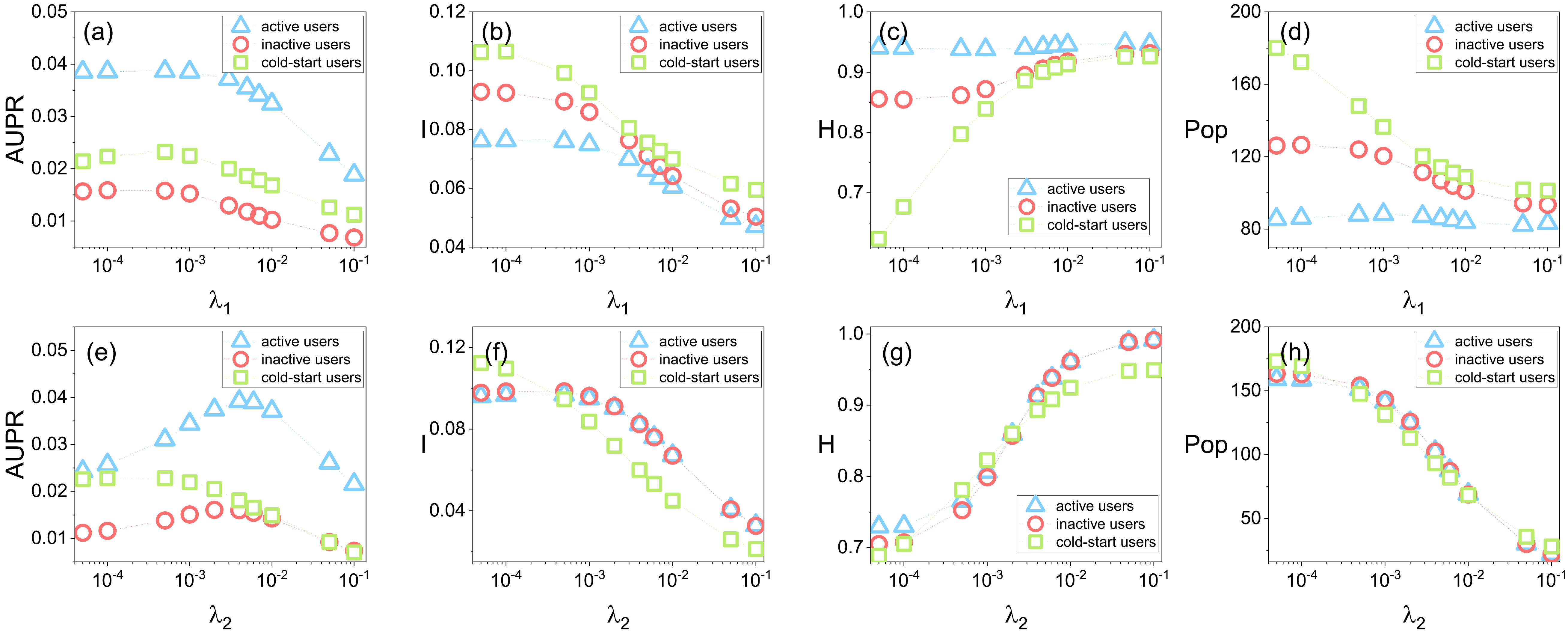}
    \caption{On Epinions, the performance of SBLO for three types of users. Except cold-start users, all results are averaged over 20 independent runs. (a)-(d) Evaluation metrics vs. $\lambda_1$, and the $\lambda_2$ values are tuned to their optimal values. (e)-(h) Evaluation metrics vs. $\lambda_2$, and the $\lambda_1$ values are also tuned to their optimal values.}
\label{fig.5}
\end{figure}

\begin{figure}[t]
\setlength{\abovecaptionskip}{0pt}
\centering
	\includegraphics[width=1\textwidth]{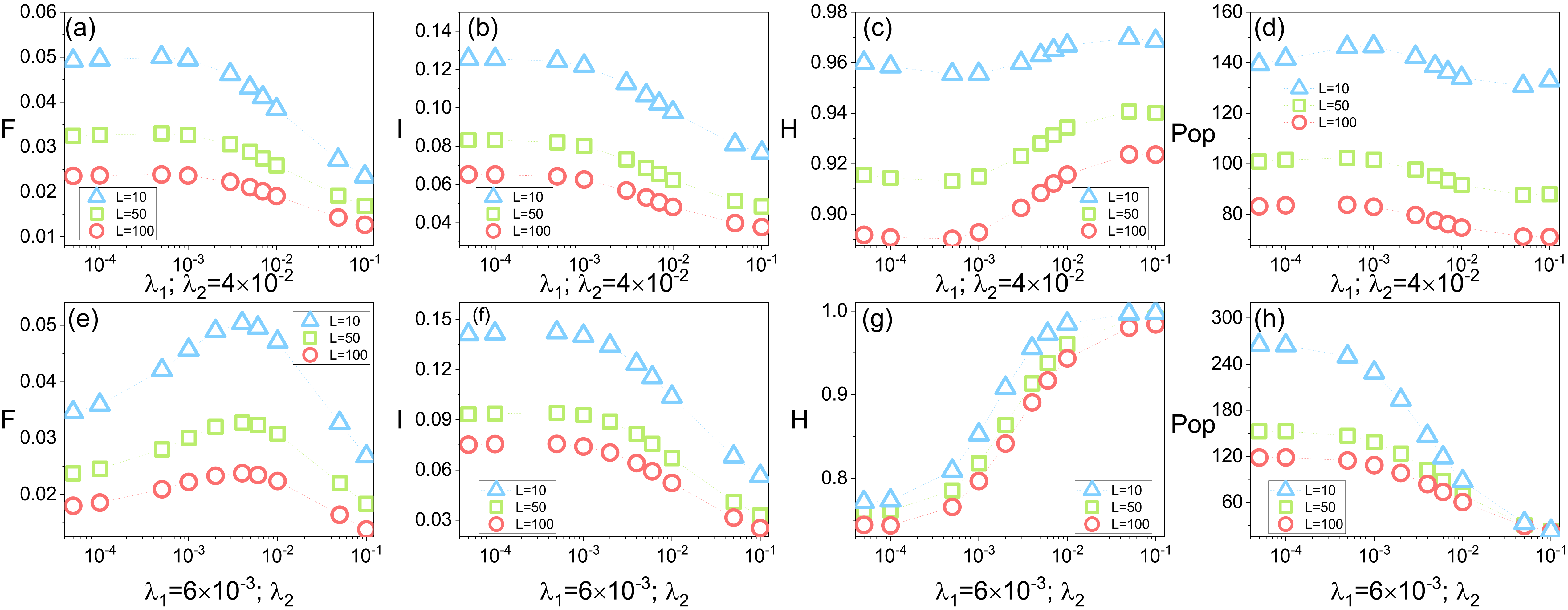}
    \caption{On Epinions, the performance of SBLO for different lengths of recommendation lists for all users. All results are averaged over 20 independent runs. (a)-(d) Evaluation metrics vs. different $\lambda_1$, $\lambda_2$=4$\times$10$^{-2}$. (e)-(h) Evaluation metrics vs. different $\lambda_2$, $\lambda_1$=6$\times$10$^{-3}$.}
\label{fig.6}
\end{figure}

Finally, we test the parameter sensitivity of SBLO in different scenarios. As shown in Fig.~\ref{fig.4}, on Epinions, for all users, SBLO is more sensitive to $\lambda_2$ than $\lambda_1$ subject to AUPR, and relatively stable subject to three diversity metrics including intra-similarity, hamming distance, and popularity. Similar findings also can be found for three types of users and different lengths of recommendation lists (see Fig.~\ref{fig.5} and Fig.~\ref{fig.6}). I.e., for users with different data sparsity or for different lengths of recommendation lists,  SBLO is more sensitive to $\lambda_2$ than $\lambda_1$ on AUPR, and relatively stable subject to three diversity metrics. Since AUPR does not dependent on recommendation lengths, we replace AUPR with F-score in Fig.~\ref{fig.6}. Another finding is that for users with different data sparsity, optimal parameters of SBLO fluctuate little, and optimal parameters for different lengths of recommendation lists also fluctuate little. Similar results also can be found on FriendFeed.

\clearpage
\section{Analysis}
\label{analysis}
There is a general phenomenon: compared with algorithms that only consider historical behaviors, algorithms that also consider social relationships can significantly improve the recommendation performance for some datasets, while for some other datasets the improvement is non-significant. This promotes us to answer how much do social relationships in SBLO contribute. 

\begin{figure}[t]
\setlength{\abovecaptionskip}{0pt}
\centering
	\includegraphics[width=1\textwidth]{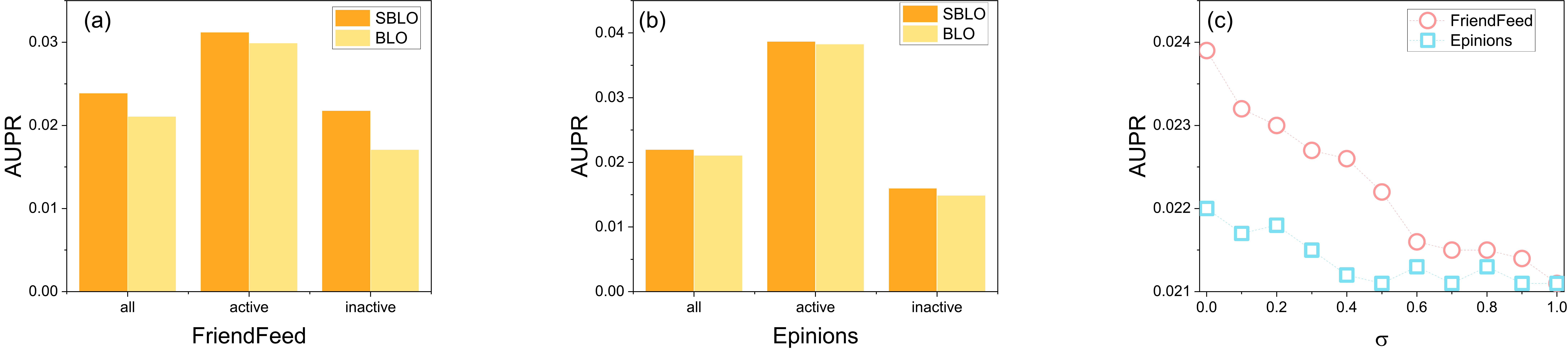}
    \caption{(a)-(b) AUPR of SBLO and BLO. (c) AUPR of SBLO vs. social perturbation ratio $\sigma$. All results are averaged over 20 independent runs.}
\label{fig.7}
\end{figure}

We first compare the recommendation performance of SBLO with its' degenerated algorithm subject to AUPR for all users. We can obtain the degenerated algorithm of SBLO if $\lambda_1$ in Eq. (\ref{eq.8}) is set to 0. In this case, the contribution of social relationships is ignored, and we name this degenerated algorithm as BLO. The corresponding score matrix is
\begin{alignat}{2}
\mathbf{R} & = \mathbf{S}^*\mathbf{B} =\lambda_2 \mathbf{B}\mathbf{B}^T(\lambda_2 \mathbf{B}\mathbf{B}^T+\mathbf{I})^{-1}\mathbf{B}.
\label{eq.22} 
\end{alignat}
Fig.~\ref{fig.7}(a) and Fig.~\ref{fig.7}(b) report AUPR of SBLO and BLO on FriendFeed and Epinions, respectively. Compared with BLO, the improvement of SBLO is different for three types of users on two datasets. On FriendFeed, the improvement of SBLO for three types of users is significant, while the improvement is relative small on Epinions. Namely, on Epinions, social relationships in SBLO contribute little to the recommendation performance. 

To further explore the contribution of social relationships in SBLO, we elaborately design a randomization model. Considering an extreme case, if relationships in social networks are randomly generated, social relationships will be meaningless to recommendation systems. So we simulate the relevance between social networks and user-object interaction networks by randomly perturbing a certain proportion of relationships in social networks. In each simulation, to decouple part of the relevance between social networks and user-object interaction networks, we perturb $\sigma$ relationships in social networks by first-order null model (with the same degree sequence obtained by link-crossing operations \cite{maslov2002specificity}). As indicated in Fig.~\ref{fig.7}(c), the improvement of SBLO on AUPR declines with the increase of $\sigma$ on both of FriendFeed and Epinions. That is, the higher the relevance, the more significant the improvement of SBLO. If there is no relevance between social networks and user-object interaction networks, social relationships in SBLO will not contribute to the recommendation performance. 

\section{Conclusion and discussion}
\label{conclusion}
Social network is a great fortune for recommendation systems. How to make better use of it remains a challenge. This paper presents an alternative solution to this challenge: through jointly constraints from social networks and user-object interaction networks, we can obtain social relationships that are valuable for recommendation systems. While social relationships will does not work if there is no relevance between social networks and user-object interaction networks. In this case, the performance enhancement of SBLO only be guaranteed by the linear optimization technique. It is also meaningful to explore the relation between implicit factors and probabilities of link formation between nodes, so as to design an objective function that is more suitable for the target dataset. Obviously, it will be very valuable in practical applications, and we leave this question for future studies. 

Overall speaking, SBLO outperforms three classical algorithms based on historical behaviors, and three representative algorithms based on both historical behaviors and social relationships subject to accuracy metrics. Moreover, SBLO also can improve the recommendation performance for inactive users and cold-start users. An unexpected finding is that SBLO is competitive with algorithms that dedicated to diversity, like PD and HHP. It is worth noting that the recommendation accuracy of CosRA+T is second only to SBLO, but it does not work for cold-start users. 

\section*{Acknowledgments}
This work was partially supported by the Key Scientific Research Fund of Xihua University under Grant No. Z222022, the Ministry of Education of Humanities and Social Science Project under Grant No. 21JZD055, and the National Natural Science Foundation of China (Grant Nos. 11975071 and 61802316). 

\bibliography{mybibfile}
\end{document}